\documentclass[11pt]{article}
\usepackage{amsmath, amssymb}

\newtheorem{proposition}{Proposition}

\def \BB  {{\cal B}}

\def \FF  {{\cal F}}

\def \CCC {\mathbb{C}}

\def \HHH {\mathbb{H}}

\def \PPP {\mathbb{P}}

\def \RRR {\mathbb{R}}

\def \ZZZ {\mathbb{Z}}

\title{Two examples of discrete-time quantum walks taking continuous steps}
\author{ Alex D. Gottlieb}
\date{}
\begin{document}
\maketitle

\begin{abstract}This note introduces some examples of quantum random walks in
$\RRR^d$ and proves the weak convergence of their rescaled
$n$-step densities.   One of the examples is called the Plancherel
quantum walk because the ``quantum coin flip" is the Fourier
Integral (or Plancherel) Transform.  The other examples are the
Birkhoff quantum walks, so named because the coin flips are
effected by means of measure preserving transformations to which
the Birkhoff's Ergodic Theorem is applied.
 \end{abstract}

 Quantum walks of the type we consider in this note
were introduced in \cite{ABNVW}, which defined and analyzed the
Hadamard quantum walk on $\ZZZ$, and a ``new type of convergence
theorem" for such quantum walks on $\ZZZ$ was discovered by Konno
\cite{K1,K2}.  A much simpler proof of Konno's theorem has
recently appeared in \cite{GJS}, allowing the theorem to be
generalized to quantum walks in $\ZZZ^d$. Inspired by the
technique of \cite{GJS}, I have proven that Konno's theorem also
holds for an analog of the quantum walk that takes steps in
$\RRR^d$ instead of $\ZZZ^d$.

In this note, I describe a couple of quantum walks that take steps
in $\RRR^d$: the Birkhoff quantum walk and the Plancherel quantum
walk.  These are analogs of the Hadamard quantum walk of
\cite{ABNVW}, which is reviewed next.

The Hadamard random walker steps along the lattice $\ZZZ$,
carrying with her a ``quantum coin."   Formally, the Walker\&Coin
state is specified by a unit vector in $\ell^2(\ZZZ)\otimes
\CCC^2$; the standard basis vectors of the auxilliary ``coin
space" $\CCC^2$ will be denoted $|H\rangle$ for ``Heads" and
$|T\rangle$ for  ``Tails."  A complete measurement of the walker's
position would find her at $j \in \ZZZ$ with probability
\begin{equation}
\label{prob}
      P(j;\psi) \ = \ \Big|\big\langle (j \otimes H) \big| \psi \big\rangle\Big|^2
      + \Big|\big\langle (j\otimes T) \big| \psi \big\rangle\Big|^2
\end{equation}
 if the state of the Walker\&Coin is $\psi \in \ell^2(\ZZZ)\otimes
\CCC^2$. But the walker walks unobserved, and her position will
become entangled with the state of her coin.  To take a step, the
quantum walker flips her coin by a Hadamard transform
\begin{eqnarray}
        |H \rangle & \longmapsto &   \tfrac{1}{\sqrt{2}}(|H\rangle + |T\rangle)   \nonumber \\
        |T \rangle & \longmapsto &   \tfrac{1}{\sqrt{2}}(|H\rangle - |T\rangle)
        \label{Hadamard}
\end{eqnarray}
and takes one step to the left or right depending on the outcome.
This conditional step is implemented by the unitary operator $S$
on $\ell^2(\ZZZ)\otimes \CCC^2$ defined by
\begin{eqnarray*}
        S(|j \rangle \otimes |H \rangle)  & = &  |j +1 \rangle \otimes |H \rangle  \\
        S(|j \rangle \otimes |T \rangle)  & = &  |j -1 \rangle \otimes |T \rangle
        \ ,
\end{eqnarray*}
so that a single step of the quantum random walk changes the
Walker\&Coin state from $\psi$ to $S(I\otimes F)\psi$, where $F$
is the Hadamard operator of (\ref{Hadamard}) and $I$ denotes the
identity operator on $\ell^2(\ZZZ)$.  Taking $n$ unobserved steps
of the Hadamard random walk changes the initial state $\psi_0$
into $U^n\psi_0$, where $U$ denotes $S(I\otimes F)$.  Konno's
theorem states that the probability measures
\begin{equation}
\label{Konno}
         \sum_{j\in \ZZZ} P(j\ ;U^n\psi_0 )\ \delta(j/n)
\end{equation}
converge weakly to a probability measure depending on $\psi_0$,
but supported in any case on the interval
$\big[\tfrac{-1}{\sqrt{2}},\tfrac{1}{\sqrt{2}}\big]$. In
(\ref{Konno}), the probabilities $P( \cdot\ ; U^n\psi_0 )$ are as
defined in (\ref{prob}) and $\delta(x)$ denotes a point-mass at
$x$.  Weak convergence $Q_n \longrightarrow Q$ of probability
measures means that $\int f dQ_n \longrightarrow \int f dQ$ for
all bounded and continuous functions $f$ on $\RRR$.

Now let us introduce a couple of analogs of Hadamard random walk
that take steps in $\RRR^d$. Instead of a quantum coin with the
alternatives $H$ and $T$, the walker will use another copy of
$\RRR^d$ to choose her next step; instead of the Hadamard
transform on $\CCC^2$, she will use a unitary operator on an
inifinite dimensional Hilbert space.  In Plancherel quantum walk,
the ``coin space" is $L^2(\RRR^d)$ and the ``coin flip" operator
is the Fourier transform on that space. (The fact that the Fourier
transform is a unitary operator on $L^2(\RRR^d)$ is known as
Plancherel's Theorem \cite{Wiener}.) In a Birkhoff quantum walk,
the coin space is $L^2(\Omega,\BB,\PPP)$ where $(\Omega,\BB,\PPP)$
is a probability space, and the coin flip operator is the unitary
map
\begin{equation}
\label{BirkhoffFlip}
        (F_T f)(\omega) \ = \ f(T(\omega))
\end{equation}
generated by a measure-preserving transformation $T$ (i.e., a
measurable map from $\Omega$ to itself, whose inverse exists and
is also measurable, and such that $\PPP(T(E)) = \PPP(E)$ for all
measurable $E \subset \Omega$ \cite{Petersen}).

In a Plancherel quantum walk, the Hilbert space for the
Walker\&Coin is $\HHH = L^2(\RRR^d)\otimes L^2(\RRR^d)$. This
space is isomorphic to $L^2(\RRR^{2d})$ and its members may be
represented by wavefunctions $\psi(x,y)$ with $x,y \in \RRR^d$.
The unitary operator $\psi \longmapsto U\psi$ on $\HHH$ with
\begin{equation}
\label{PlancherelStep}
        (U\psi)(x,y) \ = \ (2\pi)^{-d/2}  \int \psi( x-y, t)  e^{-ity}dt
\end{equation}
determines the single step of the Plancherel quantum random walk.
This is the composition of the ``conditional step" operator
\[ (S\psi)(x,y) \ = \ \psi(x - y, y) \]
with the ``coin flip" operator $I\otimes \FF$, where $\FF$ denotes
the Fourier transform on $L^2(\RRR^{d})$.

In a Birkhoff quantum walk, the Hilbert space for the Walker\&Coin
is $\HHH = L^2(\RRR^d)\otimes L^2(\Omega,\BB,\PPP)$.  This space
is isomorphic to $L^2(\RRR^d \times \Omega)$ and its members may
be represented by wavefunctions $\psi(x,\omega)$ with $x \in
\RRR^d, \omega \in \Omega$.   Let $h$ be an integrable function on
$\Omega$ with values in $\RRR^d$.   The unitary operator $\psi
\longmapsto U\psi$ on $\HHH$ with
\begin{equation}
\label{BirkhoffStep}
        (U\psi)(x,\omega) \ = \ \psi(x - h(\omega), T(\omega))
\end{equation}
determines the single step of the Birkhoff quantum random walk.
This is the composition of the conditional step operator
\[ (S\psi)(x,\omega) \ = \ \psi(x - h(\omega), \omega) \]
with the coin flip operator $I\otimes F_T$, where $F_T$ is defined
in (\ref{BirkhoffFlip}).

The following are the analogs of Konno's theorem for the Birkhoff
and Plancherel quantum walks.
\begin{proposition}
\label{Birkhoff}

Let $U$ be as in (\ref{BirkhoffStep}).
 For an arbitrary but fixed initial state $\psi_0
\in \HHH$, define the probability densities
\begin{equation}
\label{BirkhoffPn}
    P_n(x) \ = \  \int \big| U^n\psi_0(x,\omega ) \big|^2 \PPP(d\omega)
\end{equation}
on $\RRR^d$.
 Then the rescaled
probability measures $n^d P_n(n x)dx$ converge weakly as $n
\longrightarrow \infty$.  Their weak limit is the image under
\begin{equation}
\label{pointwise}
 \overline{h}(\omega) \ = \
\lim_{n\rightarrow\infty} \frac{1}{n}\sum\nolimits_{j=0}^{n-1}
h(T^{-j}(\omega)).
\end{equation}
of the probability measure on $(\Omega,\BB)$ that has density
$
     \int | \psi_0( x,\omega )|^2 dx
$
relative to $\PPP$.
\end{proposition}
\begin{proposition}
\label{Plancherel}

 Let $U$ be as in (\ref{PlancherelStep}).
 For an arbitrary but fixed initial state $\psi_0
\in \HHH$, define the probability densities
\begin{equation}
\label{Pn}
    P_n(x) \ = \  \int \big| U^n\psi_0(x,y) \big|^2 dy
\end{equation}
on $\RRR^d$.
  Then the rescaled
probability measures $Q_n(x)dx = n^d P_n(n x)dx$ converge weakly
to the probability measure with density
\[
     Q(x) \ = \ \frac{2}{(2\pi)^d}\int \Big| \int \psi_0(t,y) e^{2 i t x}dt \Big|^2 dy
\]
as $n \longrightarrow \infty$.  Note that the limiting density
$Q(x)$ is independent of $\chi_0$ if $\psi_0(x,y) =
\phi_0(x)\chi_0(y)$.
\end{proposition}
 \noindent {\bf Proof of Proposition~\ref{Birkhoff}}:
From (\ref{BirkhoffStep}),
  $      (U^n\psi_0)(x,\omega) \ = \ \psi_0\big(x - \sum_{j=0}^{n-1}
h(T^j(\omega)),\ T^n(\omega)\big) $, and the rescaled probability
density $n^d P_n(n x)$ is
\[
        n^d \int \Big|
        \psi_0\big(nx - \sum\nolimits_{j=0}^{n-1} h(T^j(\omega)),\ T^n(\omega)\big)
        \Big|^2 \PPP(d\omega)\ .
\]
For any test function $\phi(x)\in C_b(\RRR^d)$,
\begin{eqnarray}
        \big\langle n^d P_n(n x)dx,\ \phi(x) \big\rangle
        & = &  n^d \int \phi(x) \int \Big|
        \psi_0\big(nx - \sum\nolimits_{j=0}^{n-1} h(T^j(\omega)),\ T^n(\omega)\big)
        \Big|^2 \PPP(d\omega)  dx \nonumber \\
         & = &   \int \int \phi\Big(\tfrac{1}{n}y + \tfrac{1}{n}\sum\nolimits_{j=0}^{n-1}
h(T^j(\omega)) \Big) \big|
        \psi_0\big(y,\ T^n(\omega)\big)
        \big|^2 \PPP(d\omega)  dy \nonumber \\
         & = &   \int \int \phi\Big(\tfrac{1}{n}y + \tfrac{1}{n}\sum\nolimits_{j=1}^{n}
h(T^{-j}(\omega')) \Big) \big|
        \psi_0\big(y,\ \omega' \big)
        \big|^2 \PPP(d\omega')  dy \label{changed}
\end{eqnarray}
making the changes of variables $y = nx - \sum_{j=0}^{n-1}
h(T^j(\omega))$ and $\omega' = T^n(\omega)$.  By Birkhoff's
Ergodic Theorem, the limit (\ref{pointwise}) exist almost
everywhere and defines an integrable function.  Applying the
Dominated Convergence Theorem to (\ref{changed}) yields
\[
\lim_{n\rightarrow\infty} \big\langle n^d P_n(n x)dx,\ \phi(x)
\big\rangle \ = \ \int \int \phi( \overline{h}(\omega')) \big|
        \psi_0\big(y,\ \omega' \big)
        \big|^2 dy \PPP(d\omega')\ ,
\]
which shows that $n^d P_n(n x)dx$ converges weakly to the
probability measure described in the theorem.
 \hfill
$\square$

 \noindent {\bf Proof of Proposition~\ref{Plancherel}}:  The
plan of the proof is to show that the Fourier transforms of the
probability densities $Q_n$ converge pointwise to the Fourier
transform of the probability density $Q$, for this would imply
that $Q_n(x)dx$ converges weakly to $Q(x)dx$. We denote the
Fourier transform on $L^2(\RRR^{d})$ by $\FF$, and we also define
two unitary operators $\FF_1$ and $\FF_2$ on $L^2(\RRR^{2d})$ by
\begin{eqnarray*}
        (\FF_1 f)(\zeta,y) & = &  (2\pi)^{-d/2} \int f(x,y) e^{-i x
        \cdot \zeta} dx \\
        (\FF_2 f)(x,\zeta) & = &  (2\pi)^{-d/2} \int f(x,y) e^{-i y
        \cdot \zeta} dy\ .
\end{eqnarray*}

\noindent \underline{Step 1}:\qquad For the first step we will
assume that $(\FF_1\psi_0)(\zeta,y)$ is bounded and continuous,
and we will prove that $\FF Q_{4m}$ converges to $\FF Q$ as $m
\longrightarrow \infty$.

 Since the integrand in
(\ref{Pn}) is the square of the modulus of $U^n\psi_0$, the
Fourier transform of $P_{n}(x)$ is
\begin{equation}
\label{FourierTransfomOfPn}
        (2\pi)^{-d/2} \int \int \overline{(\FF_1 U^{n}\psi_0)(\eta,y)}
                     (\FF_1 U^{n}\psi_0)(\eta + \zeta,y) d\eta dy
\end{equation}
 Let $\widetilde{U} = \FF_1 U \FF_1^*$.
It may be verified that
\[
       ( \widetilde{U}^4 \phi) (\zeta,y) \ = \ e^{i\zeta^2}
       \phi(\zeta,y)
\]
(to this end it may be helpful to note that $\widetilde{U} = M
\FF_2$ where $M$ denotes the multiplication operator $M
\phi(\zeta,y)= e^{-i\zeta y} \phi$).
 It follows that
\[
    (\FF_1 U^{4m}\psi_0)(\zeta,y) \ = \ (\widetilde{U}^{4m} \FF_1\psi_0)(\zeta,y)
                \ = \ e^{im\zeta^2}(\FF_1\psi_0)(\zeta,y) .
\]
Substituting this into (\ref{FourierTransfomOfPn}) shows that
\begin{eqnarray*}
      (\FF P_{4m})(\zeta) & = &
       (2\pi)^{-d/2} \int \int       e^{-im\eta^2}
       \overline{(\FF_1\psi_0)(\eta,y)}
        e^{im(\eta + \zeta)^2} (\FF_1 \psi_0)(\eta + \zeta,y) d\eta
        dy \\
        & = &
       (2\pi)^{-d/2} \int \int e^{im (2\eta \zeta + \zeta^2)}
       \overline{(\FF_1\psi_0)(\eta,y)}
        (\FF_1 \psi_0)(\eta + \zeta,y) d\eta
        dy
\end{eqnarray*}
and therefore the Fourier transform of the rescaled density $n^d
P_n(n x)$ is
\begin{equation}
\label{FourierTransform}
  (2\pi)^{-d/2} \int \int e^{im (2\eta \zeta/n +  (\zeta/n)^2)}
       \overline{(\FF_1\psi_0)(\eta,y)}
        (\FF_1 \psi_0)(\eta + \zeta/n,y) d\eta dy
\end{equation}
when $n=4m$.  The integrand in (\ref{FourierTransform}) tends
pointwise to $e^{i \eta \zeta / 2}|(\FF_1 \psi_0)(\eta,y)|^2$ as
$m \longrightarrow \infty$.  If $\FF_1 \psi_0$ is both bounded and
integrable for a.e. $y$, then the Dominated Convergence Theorem
implies that
\[
  (\FF Q_n)(\zeta) \ \longrightarrow \ (2\pi)^{-d/2}
  \int \int e^{i \eta \zeta / 2} \big|(\FF_1 \psi_0)(\eta,y)\big|^2 dy d\eta
\]
by the Dominated Convergence Theorem.  But the latter is the
Fourier transform of
\begin{equation}
\label{Q}
     Q(x) \ = \ 2\int \big|(\FF_1 \psi_0)(-2x,y)\big|^2 dy.
\end{equation}
This proves that the probability measures
\begin{equation}
\label{Qn}
     Q_n(x)dx \ = \ n^d P_n(n x)dx
\end{equation}
converge weakly to $Q(x)dx$ along the subsequence $n=4m$ as $m
\longrightarrow \infty$.

\noindent \underline{Step 2}:\qquad Next, a density argument
removes the restriction on $\psi_0$ in Step~1:

Let $\psi_0^{(j)}$ be a sequence of normalized wavefunctions that
converges to an arbitrary $\psi_0 \in L^2(\RRR^{2d})$. The
$\psi_0^{(j)}$ may be chosen from the Schwartz class, which is
dense in $L^2(\RRR^{2d})$.  By Step~1, the probability measures
$Q_{4m}^{(j)}(x)dx$ converge weakly to $Q^{(j)}(x)dx$ as $m
\longrightarrow \infty$, where $Q_{4m}^{(j)}$ and $Q^{(j)}$ are
defined as in (\ref{Q}) and (\ref{Qn}) with $\psi_0^{(j)}$ in
place of $\psi_0$.   On the other hand, the Cauchy-Schwartz
inequality implies that $    \big\|Q^{(j)} - Q\big\|_1
    =  2 \| \psi_0^{(j)} - \psi_0 \|_2  $ and indeed
\[
    \big\|Q_{4m}^{(j)} - Q_{4m}\big\|_1
    \  \le \ 2 \big\| \psi_0^{(j)} - \psi_0 \big\|_2
\]
for all $m$.  The weak convergence $ Q_{4m}^{(j)} dx
\longrightarrow Q^{(j)} dx $
 and the preceding
uniform bound on $\|Q_{4m}^{(j)} - Q_{4m}\|_1$ imply that
$Q_{4m}dx$ convergences weakly to $Q dx$ weakly.

\noindent \underline{Step 3}:\qquad Finally, we will prove that
$Q_n$ tends to $Q$ along all subsequences, having already shown
that $Q_{4m}(x)dx \longrightarrow Q(x)dx$ for any initial state
$\psi_0$. It will help to have the notation for $P_n$ and $Q_n$
display the dependence on the initial state; from now on we will
write $P_n(x\ ;\psi)$ and $Q_n(x \ ;\psi)$ to indicate this
dependence. From (\ref{Pn}) and (\ref{Qn}) one has that
\begin{eqnarray}
\Big\| Q_{n+p}( x \ ;\psi_0) & - & Q_n( x \ ;U^p\psi_0) \Big\|_1
\label{difference} \\
& = &
\int \Big| (n+p)^d P_{n+p}((n+p)x\ ;\psi_0) - n^d P_n(nx\ ;U^p\psi_0) \Big| dx \nonumber \\
& = & \int \Big| (1+\tfrac{p}{n})^d P_n\left((1+\tfrac{p}{n})u\
;U^p\psi_0\right) - P_n(u\ ;U^p\psi_0) \Big| du
 \nonumber
\end{eqnarray}
for any positive integer $p$, and therefore $\| Q_{n+p}( x \
;\psi_0) - Q_n( x \ ;U^p\psi_0) \|_1$ tends to $0$ as $n
\longrightarrow \infty$ for fixed $p$ since translation acts
continuously on $L^1$. Steps~1 and 2 of this proof and the
estimate (\ref{difference}) imply that
\[
Q( x \ ;U^p\psi_0)dx \ = \ \lim_{m\rightarrow \infty } Q_{4m}( x \
;U^p\psi_0)dx \ = \ \lim_{m\rightarrow \infty } Q_{4m+p}( x \
;\psi_0)dx\ .
\]
On the other hand, one may show by induction that $Q( x \
;U^p\psi_0) = Q( x \ ;\psi_0)$ for all $p$.  It follows that
$
     Q_n(x\ ;\psi_0)dx  \longrightarrow Q(x\ ;\psi_0)dx
$ weakly along {\it every} subsequence.  \hfill $\square$


\begin{thebibliography}{X}


\bibitem{ABNVW} A. Ambainis, E. Bach, A. Nayak, A. Vishwanath, and
J. Watrous. One-dimensional quantum walks.  {\it Proceedings of
STOC'01},\  37-49 (2001)

\bibitem{Dudley}  R. Dudley.  {\it Real Analysis and Probability}.
 Cambridge University Press, Cambridge, UK (2002)

\bibitem{GJS} G. Grimmett, S. Janson, P. Scudo.  Weak limits for
quantum random walks.  Preprint: \  quant-ph/0309135 (2003)


\bibitem{K1} N. Konno.  Quantum random walks in one dimension.
{\it Quantum Information Processing}\ {\bf 1}: 345-354 (2002)


\bibitem{K2} N. Konno.  A new type of limit theorems for the
one-dimensional quantum random walk.  Preprint: \quad
quant-ph/0206103 (2002)

\bibitem{Petersen} K. Petersen.  {\it Ergodic Theory}.
Cambridge University Press (1983)

\bibitem{Wiener} N. Wiener.  {\it The Fourier Integral and Certain of Its Applications}.
Dover Publications, New York (1933)

\end{thebibliography}
\end{document}